\documentclass[conference]{IEEEtran}
\usepackage{graphicx}
\usepackage{url}
\usepackage{caption}
\usepackage{longtable}
\usepackage{multirow}
\usepackage{pbox}
\usepackage{url}
\usepackage{lscape}
\usepackage{verbatim} 
\usepackage{array}
\usepackage{adjustbox,lipsum}
\usepackage[ruled,vlined,linesnumbered]{algorithm2e}
\usepackage{caption}
\usepackage{footnote}

\begin{document}

\title{Application of Case-Based Teaching and Learning in Compiler Design Course}

\author{\IEEEauthorblockN{Divya Kundra}
\IEEEauthorblockA{Deen Dayal Upadhyaya, India\\
divya1395@iiitd.ac.in}
\and
\IEEEauthorblockN{Ashish Sureka}
\IEEEauthorblockA{ABB Corporate Research, India\\
ashish.sureka@in.abb.com}}
\maketitle
\begin{abstract}
Compiler design is a course that discusses ideas used in construction of programming language compilers. Students learn how a program written in high level programming language and designed for humans understanding is systematically converted into low level assembly language understood by machines. We propose and implement a Case-based and Project-based Learning environment for teaching important Compiler design concepts (CPLC) to B.Tech third year students of a Delhi University (India) college. A case is a text that describes a real-life situation providing information but not solution. Previous research shows that case-based teaching helps students to apply the principles discussed in the class for solving complex practical problems. We divide one main project into sub-projects to give to students in order to enhance their practical experience of designing a compiler. To measure the effectiveness of case-based discussions, students complete a survey on their perceptions of benefits of case-based learning. The survey is analyzed using frequency distribution and chi square test of association. The results of the survey show that case-based teaching of compiler concepts does enhance students skills of learning, critical thinking, engagement, communication skills and team work.
\end{abstract}

\IEEEpeerreviewmaketitle

\section{Research Motivation and Aim}

A majority of engineering classes involve traditional lecture-based approach in which learning is considered as oriented from teachers to students. The traditional teaching is concerned with teacher being the active controller, having the entire power and responsibility of the environment \cite{Altman2010}. The only activity on behalf of students is answering the questions posed by the teacher. The lesson's content and delivery are considered to be most important and students master knowledge through drill and practice such as rote learning \cite{theroux2004real}. This lecture-based approach is not motivating and does not prepare engineering students well for professional world. 
Seymour and Hewitt \cite{seymour1997talking} reported that use of traditional methods of teaching has led to low level of attendance and retention in engineering disciplines. The use of case-based pedagogy can offer solutions to prepare students for the professional world, make education motivating and reduce attrition rates. Case-based learning is different from traditional learning in the manner that it places students as the center of education process. Students are given importance in what and how they are learning. Cases can be problem-based, historical in nature, present a model, dilemma-based or demonstrate critical issues in the field \cite{yadav2009using}. Students apply the theoretical knowledge in solving practical world problems in a supportive environment \cite{INSP:INSP002}. Real world problems are usually complex, ill-structured, have conflicting choices and can be presented in number of ways to students \cite{jonassen2006everyday}. 

Application of case-based learning is useful in learning about compiler design concepts for the following reasons :

\begin{enumerate}

\item \textbf{Making learning easier and interesting} \cite{aa}: Compiler design course has conceptually difficult topics. It is not easy to teach particularly in small college environment. There are insufficient small grammar examples supported by main textbooks while in reality the grammars for commonly used languages are too complex. Thus use of contemporary approaches like case-based can enhance the understanding of the course while keeping the class engaging.

\item \textbf{Understanding implementation of real world software} \cite{Jonassen2002}: We develop cases from real world software which use the core concepts of compiler design. We believe students can understand and practice how things actually work in real world.

\item \textbf{Skill building} \cite{kaddoura2011critical}: Through repeated exposure to ambiguous and complex problems in cases, students build confidence and critical thinking. It exposes them to ambiguities and enhances abilities to take timely and effective decisions to unclear and complex problems.

\item \textbf{Addition to case repository} \cite{shankararamanteaching}: To our best of the knowledge, not much work has been done in teaching compiler design using case-based teaching methodology. Thus cases developed can be shared and used by other faculty while teaching compiler design course.

\vspace*{1\baselineskip}

Compiler design course involves element of programming. Writing a compiler by self can give students experience of large scale application development. Thus programming projects needs to be included in the course contents. Hence, we give mini projects at the completion of two major phases of the compiler- lexical and syntax analysis. The research aims of the work presented in the paper are as following:

\begin{enumerate} 
   \item  Develop cases for teaching essential concepts of compiler design.    \item Propose a complete teaching framework that teaches important concepts of compiler design using case-based and project-based learning approaches.
    \item Investigate the effectiveness of case discussions.
\end{enumerate}

\section{Related Work and Research Contributions}
Literature shows the use of case-based learning for teaching computer science courses. In \cite{liu2014application} the authors adapt the case-based learning method during the teaching of Delphi language for teaching object oriented concepts. In \cite{he2013supporting} authors presents a case study to show value of case-based learning in improving the teaching of information security. Authors in \cite{gargcase} have used case studies to teach software engineering. There have been use of different techniques to teach compiler design course. In \cite{Shapiro:1976:NAT:953026.803467} authors present an approach in which the traditional term project is replaced by several small independent assignments. Each assignment itself is a small compiler of a small programming language which is to be built using different parsing techniques. \cite{Vegdahl:2000:UVT:369340.369325} uses visualization tools to teach compilers. Author uses a pair of packages that employ Java's graphical capabilities so that a program can be visualized at various stages of compilation process. \cite{aa} proposes effective approaches in teaching principles of compiler that includes concept mapping, problem-based learning (PBL), case study and e-learning.
 \par
\setlength{\parindent}{5ex}
Some tools have also been built to facilitate learning of compilers. In \cite{1183668} authors have build a tool LISA that supports learning and conceptual understanding of compiler design in an efficient, direct and long lasting manner. \cite{Demaille:2008:STT:1384271.1384291} introduces a set of tools designed and improved for compiler design educative projects in C++. In \cite{Xu:2006:CCT:1124706.1121370} \textit{Chirp}- a language specification and compiler implementation is proposed. As a language \textit{Chirp} is matched with stack-based virtual machine that is build on simple handy cricket educational robot controller. As a compiler \textit{Chirp} is designed into series of components with each component demonstrating compiler construction technique. Compiler construction has also been taught through domain specific language. \cite{Henry:2005:TCC:1047124.1047364} suggests that building a compiler for domain specific language (language specially designed for some specific problem) can engage students more than traditional compiler projects. In \cite{Ruckert:2007:TCC:1227504.1227460} authors argue that compiler teaching through an unusual programming language \textit{textttklx} with target processor as the postscript interpreter is a good choice for teaching compilers. Work in \cite{Mallozzi:2005:TTT:1089053.1089078} describes an approach using tools developed by the author to generate a parser that encourages learning of object-oriented techniques. \cite{White:2005:HSU:1047124.1047365} presents an idea of integrating real compiler code into teaching theory of compilers. 
Authors use debugger on compiler in directed ways resulting in students being shown the relevant parts of compilers internals. 

\vspace*{1\baselineskip}

In context to previous existing work, our paper makes following novel research contributions :
\begin{enumerate}
    \item We developed cases for teaching core compiler concepts. 
     \item We propose and implement a complete teaching framework CPLC which contains learning objectives, case-based and project-based pedagogy and measures students understanding.  
     \item We evaluate the performance and impact of the proposed case-based pedagogy and demonstrate its effectiveness.
\end{enumerate}

\noindent \textbf{Dataset Contributions}: We make our dataset publicly available on GitHub\footnote{\url{https://github.com/Divya-Kundra/Case-Based-Teaching}}. GitHub is becoming popular as a platform for researchers and scientists to share, update and maintain their dataset as well as code\footnote{\url{http://www.nature.com/news/democratic-databases-science-on-github-1.20719}}. We believe that sharing our dataset will further facilitate research on case-based teaching in computer science and in particular on compilers design course and can be used to explore new research problems and hypothesis. Due to limited space in the paper, we briefly describe only two case-studies, however, we make all the case-studies publicly available through the GitHub repository. 
\\\newline
\noindent \textbf{Extended Version of Previous Paper}: The study presented in this paper is an extended version of our short paper accepted in T4E 2016 (The Eighth IEEE International Conference on Technology for Education) by the same authors \cite{kundra2016}. Due to the limited four page limit of the T4E 2016 paper, several aspects of our study are not covered which are now described in detail in this paper. The objective of this paper is to provide a complete and detailed analysis of our work through arXiv open access\footnote{\url{ https://arxiv.org/}}. 

\section{Learning Framework: CPLC}

We propose a learning framework CPLC- Case-based and Project-based Learning environment for teaching Compiler design concepts. We apply CPLC while teaching the analysis phase of compiler i.e. lexical and syntax analysis as the content in these phases is very important and is sufficient for learning many of the basic principles of a compiler. The model uses cases that are based on practical problems giving students hands on experience in solving complex real world problems. The model also involves giving a mini project at the conclusion of a case discussion. Projects in compiler design course serve two purposes \cite{aiken1996cool} : they help in better understanding of the language whose compiler is to be implemented and it gives students experience of building sustainable software. Projects for designing partial compilers using LEX (Lexical Analyser) and YACC (Yet Another Compiler Compiler) \cite{levine1992lex} for different languages like C, Java, HTML, SQL MATLAB and Python are given to students.     
 \par
\setlength{\parindent}{5ex}
Our proposed model CPLC is complete in itself providing learning objectives, a teaching methodology and evaluation of students understanding. The model is based upon the constructivist perspective of learning. Constructivism means learners are more actively involved in process of learning rather than just passively receiving information from teacher \cite{gray1997contructivist}. According to Audrey Gray \cite{gray1997contructivist} the characteristics of a constructivist classroom includes that learners are actively involved in a classroom and the activities that happen in the class are interactive and student-centric. The teacher facilitates the process of learning in which students are more autonomous and responsible. CPLC learning model is combination of the following pedagogical models:

\begin{figure*}
\centering
\noindent\includegraphics[width=16cm,height=8cm]{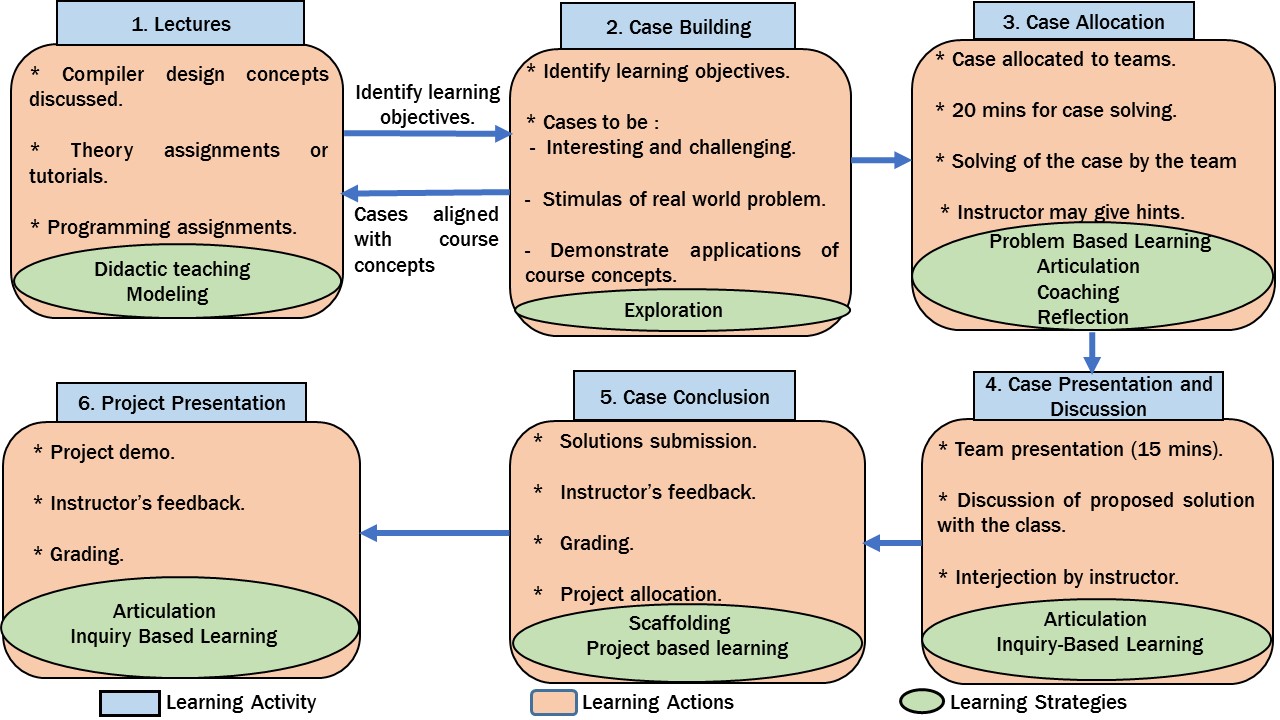}
\caption{Learning Cycle showing Learning Activity, Learning Actions and Learning Strategies.}
\label{fig:framework}
\end{figure*}
   
\begin{enumerate}
\item \textbf{Didactic Method}: In didactic method, teachers give instruction to students who are usually the passive listeners. \cite{Altman2010} explains didactic teaching as teacher oriented in which the teacher is the controller of learning environment. Students are just part of traditionally planned lessons. Entire power and responsibility lies in hands of teacher. \cite{Altman2010} mentions that students are just knowledge holes who needs to be filled with information.

\item \textbf{Problem-Based Learning}: It suggests that if students learn by solving problem, they learn both the content and thinking strategies \cite{Hmelo-Silver2004}. It is self-reflexive such that learners monitor their own understanding and learn to adjust strategies for learning \cite{hung2008problem}. Teachers are not just knowledge disseminators rather they are facilitators who support and model reasoning processes, facilitate group processes and interpersonal dynamics, probe student's knowledge deeply without providing direct answers \cite{hung2008problem}.

\item \textbf{Cognitive Apprenticeship Model}: Cognitive apprenticeship is a model in which learners learn from an expert by cognitive and metacoagnitive skills and processes \cite{dennen2008cognitive}. Its basic model consists of following teaching strategies:
\begin{enumerate}
     
\item Modelling \cite{dennen2008cognitive}- It demonstrates the thinking process by the teacher so that student can experience and build a conceptual model of the task. 

\item Coaching \cite{dennen2008cognitive}- It involves observing student's performance and giving necessary hints for improvement. 

\item Reflection \cite{dennen2008cognitive}- It includes self analysis and assessment by students on their own. The goal of reflection for students is to analyze their own performance with the objective of better understanding and improvement.

\item Exploration \cite{dennen2008cognitive}- It involves forming and testing one's own hypothesis in pursuit of learning. The subjects are given their own space to understand, solve the problem and come with an appropriate solution. It encourages them to develop interesting problems on their own and consequently develop solutions for them.

\item Scaffolding \cite{dennen2008cognitive}: Teachers provide support to the students so that they reach to a higher level of comprehension and skill acquisition which they would not be able to reach without the guidance.

\item Articulation \cite{dennen2008cognitive}: This involves verbalising the results of reflective acts. It involves human-human interaction which demonstrates knowledge and thinking process in order to expose and clarify them \cite{mclellan1994situated}.
\end{enumerate}

\item \textbf{Project-Based Learning}: This learning model organizes learning around the project \cite{thomas2000review}. Definition of this model in the literature includes that projects are challenging and complex, involve students in designing, problem-solving and decision making \cite{jones1997real}. Students work on it for extended period of time and it ends with some useful product or presentations. Some other features of this model include authentic content, authentic assessment, hints by teacher, well defined education goals cooperative learning and reflection \cite{diehl1999project}\cite{mittal2014}\cite{sripada2015}\cite{shukla2012}.    

\end{enumerate}

\par
We propose and apply the framework shown in Figure \ref{fig:framework} for teaching the lexical and syntax analysis phase of compiler design. The learning cycle includes several learning activities like lectures, case building, case allocation, case solving, case discussion, case presentation, project allocation and project presentation. Each learning activity has associated learning actions which result in learning done through different learning strategies. The foremost learning activity consists of didactic teaching in form of classroom lectures which introduces and discusses concepts of compiler designing. The instructor models solving of the problems by writing, presenting explicitly and thinking aloud. To enhance the understanding, theory and programming assignments are given to students. For the concepts discussed in class, instructor explores new ideas and viewpoints and finds analogies in real life to develop interesting cases. The learning objectives of the concepts should be challenged in the cases. Cases should deal with interesting practical world problems so that they relate to the audience and awake their interest to solve them \cite{herreid2002makes}. Once a concept is covered in the class, we allocate the case to a team of $3$-$4$ students. Multiple teams are given the same case. Evaluation guidelines are also discussed. 
\par
\setlength{\parindent}{5ex}
Once the case is assigned, students retrieve similar problems and concepts taught in class to do problem solving \cite{Jonassen2002}. Articulation through discussions within a team help students to consider different point of views, understand problem better and come up with the best solution. Instructor coaches students by monitoring their activities, assisting and supporting them whenever necessary \cite{dennencognitive}. Student reflects over his performance by self analysis and self assessment \cite{dennen2008cognitive}. The results of reflection are put into verbal form in form of presentations. Students are encouraged to create and ask inquiries resulting in inquiry-based learning. The entire class works together by discussions and debates to reach final solution.  Once the presentation is over, team submits solutions in form of a written report to the instructor. Instructor provides in depth analysis of the team's performance. Instructive feedback incorporates extra information and improvement in response to student's work \cite{van1980learning}. A suitable grade for team's presentation is awarded by the instructor. After completion of the case, a project is assigned to the team. Students decide how to approach the problem and what suitable actions to take. Scaffolding happens where the instructor guides and advises students about the projects so that they can cope up with the task situation \cite{dennencognitive}. At the end project presentation takes place where students demonstrate their results. Inquires are raised again to the presenting team for further increase in development of knowledge or solution. It is concluded by instructor's feedback which provides clear guidance on improving learning and necessary grading is done.

\subsection*{Sample Case Studies}

We developed multiple cases for lexical and syntax analysis phase. The cases have been  publically shared\footnote{https://github.com/Divya-Kundra/Case-Based-Teaching.git} so that they can be used by other instructors of compiler course. The learning objectives of lexical analysis phase include understanding tokenization, how it takes place, learning and practicing how to take decisions for building the suitable tokenizer according to requirements of the given system. Similarly for syntax analysis phase the learning objective is to understand parsing, how the parsing of the sentence takes place, learning how the parser is constructed and understanding different parsing techniques. Student should be able to correctly judge how to build a suitable grammar and parser for the system under consideration. We present summary of one of the case of lexical and syntax analysis.

\vspace*{1\baselineskip}

\subsection{\textbf{Case of Spam Detection (Lexical Analysis)} }
\textit{Developers of an upcoming email service - mails.com want to make a spam filter that automatically detects and removes spam. The filter would consists of thousands of pre-defined ‘spam-rules’ against which the email content will be compared. Anything matching to the ‘spam-rules’ would categorize to be a spam component. The developers know that as spam filters evolves to better classify spam, the spammers will adapt their writing methods to avoid detection. Thus to build effective rules, the developers of mails.com begin to observe what kind of spam attacks can occur on filters. Example as statistical spam filters begins to learn that word like ``offer" mostly occur in spam and starts to think ``offer" as spam-rule, spammers began to obfuscate them with punctuation, such as ``o.f.f.e.r". Some of the other attacks are also explained in the case. Observing the attacks discussed in the case and reasoning what other attacks can occur, appropriate tokenization mechanism is to be decided to achieve maximum accuracy of the filter.} 

\vspace*{1\baselineskip}
The challenges for the students in this case are:
    \begin{enumerate}
        \item Identify various tokenization attacks that can occur on spam filter.
        \item Analyze and describe why and how a particular attack can occur.
        \item  Decide the most promising tokenization techniques that can be proposed for the system.
        \item Evaluate the reliability of the proposed tokenization scheme by proving how it will be resilient to the attacks.
        
    \end{enumerate}
     
 \vspace*{1\baselineskip}

The teams analysed the case and presented variety of solutions for the attacks they could identify and synthesize. Some teams argued that tokenization attacks which include splitting or modifying key word features (using more of capitalisation or punctuations within the word) are most common and thus proposed solutions for them. Some presented obfuscation attacks (changing spelling of spam words to avoid detection) to be a major spam content and gave solutions for it. A few teams presented statistical errors such as adding random good words to spam or concatenating of small illegitimate words to form a big permissible word. Teams also discussed about obfuscation of URLs done by encoding or adding unnecessary parenthesis to avoid rule based detection. For data pre-processing different ideas were suggested. Many of them were to filter out stop words like is, an, the and special characters like (), [], performing word stemming and converting all letters into lower case.
\par
\setlength{\parindent}{5ex}
To counter tokenization attacks strategies suggested included scanning the content twice,  in the first scan removal of extra spaces, punctuations within the words, and in the next scan matching of each token against bag of spam words (keyword searching). Deterministic Finite Automatas were drawn by students for the keywords/spams. Some suggested count of punctuations to be an indicator of spam. Idea to use n-grams approach which takes advantage of contextual phrase information (e.g. ``buy now") was also proposed. For statistical errors different solutions presented were: keeping a count on good words to match against a threshold, weighing the good words against spam words (a significant presence of both can indicate spam) and keeping a count of location of occurrence of good words as some argued that spammers usually insert good words in the beginning or at the end. For composite attacks, ideas mentioned were use of prefix detection to detect spam by demonstrating the use of REJECT\footnote{http://dinosaur.compilertools.net/lex/} construct of YACC as done in the class. For invalid URL, suggestions to do various forms of normalisation of URL were discussed and for spam present in attachments like images, discussions were done to process the image to extract set of tokens from properties of image. Thus the case helped students to contemplate over different tokenization strategies and gain an experience on how crucial it is to design correct tokenization scheme in the real world design of a spam filter.

\begin{table*}[t]
\caption{{Students Response to Survey Questions.} }
\label{survey}
 \centering

\begin{tabular}{|m{1.5 cm}| p { 4cm} |>{\centering\arraybackslash} m{2 cm} | >{\centering\arraybackslash} m{1.5 cm}|>{\centering\arraybackslash} m{1.5 cm}| >{\centering\arraybackslash} m{2 cm} |}
 
 \hline
 \multicolumn{1}{|c|}{\textbf{{\footnotesize Learning Principles}}} &
  \multicolumn{1}{|c|}{\textbf{{\footnotesize Question }}} & \multicolumn{1}{|c|}{\textbf{{\footnotesize Strongly Agree (\%)}}} &
    \multicolumn{1}{|c|}{\textbf{{\footnotesize Agree (\%) }}} &
      \multicolumn{1}{|c|}{\textbf{{\footnotesize Disagree (\%) }}} &
        \multicolumn{1}{|c|}{\textbf{{\footnotesize Strongly Disagree (\%) }}} \\ 
 \hline\hline
 \textbf{Learning} & {I felt the use of case-based learning was relevant in learning about course concepts.} & {$14.5$ } & {$68.7$ } & {$10.4$} & {$6.25$} \\ 
 \hline
 \textbf{Learning} & The case-based learning allowed for a deeper understanding of course concepts. & $10.4$ & $58.3$ & $27.0$ & $4.1$ \\
 \hline
 \textbf{Learning} & The case study will help me to retain different aspects of compilers better.& $12.5$ & $52.0$ & $25.0$ & $10.4$\\
 \hline
 \textbf{Critical Thinking} & The case study allowed me to view an issue from multiple perspectives. & $22.9$ & $64.5$ & $8.3$ & $4.1$ \\
 \hline
 \textbf{Critical Thinking} & The case study was helpful in synthesizing ideas and information presented in course. & $12.5$ & $77.0$ & $6.25$ & $4.1$ \\
 \hline
  \textbf{Critical Thinking} & The case study added a lot of realism to class. & $14.5$ & $64.5$ & $14.5$ & $6.25$ \\ 
 \hline

 \textbf{Engagement} & I was more engaged in class during case study. & $27.0$ & $62.5$ & $4.1$ & $6.25$\\
 \hline
\textbf{ Engagement} & The case discussion increased my interests in learning about compilers. & $18.75$ & $58.3$ & $14.5$ & $8.3$\\
 \hline

\textbf{Communication Skills} & The case discussion strengthen my communication skills to speak in front of audience. & $27.0$ & $58.3$ & $12.5$ & $4.1$\\
 \hline

\textbf{Team Work} & The case discussion increased my confidence to work in a team. & $22.9$ & $62.5$ & $12.5$ & $2.0$\\
 \hline

 \hline 
\end{tabular}
\end{table*}

\vspace*{1\baselineskip}
\subsection{\textbf{Case  of Human-Robot Chess play (Syntax Analysis) } }
\textit{GOLEMS\footnote{http://www.golems.org/projects/krang.html} is a humanoid robotics lab at Georgia Institute of Technology. The lab works towards developing robots having human and even super human capabilities. One of the tasks of the lab is working on  building a physical human-robot chess. One side of the chess would have a movable robot arm with sensors providing suitable force to locate, pick, drop and rotate the chess pieces while on other side would be the human playing against the robot. The required objectives of the robot is explained in the case. Developers have come up with controlling of the robot using context-free grammars which they have called as motion grammar. The production rules of the grammar represent a task decomposition of robotic behavior. The motion grammar enables robots to handle uncertainty in the outcomes of control actions through on-line parsing. The main task is to identify various challenges that will come in design of robot human chessplay system and address those challenges by building the suitable grammar. Thus after understanding the requirements and constraints of the system  students are required to suggest a promising motion grammar.}

\vspace*{1\baselineskip}
The challenges presented to students in this case are:
     \begin{enumerate}
         \item Identify various requirements of the system to build human-robot chessplay.
         \item Identify implicit problems and factors that influence the requirements.
         \item Decide and justify the best suitable grammar that can be build which incorporates the requirements of system.
     \end{enumerate}
     
 \end{enumerate}

\begin{figure*}
\centering
\noindent\includegraphics[width=16cm,height=8cm]{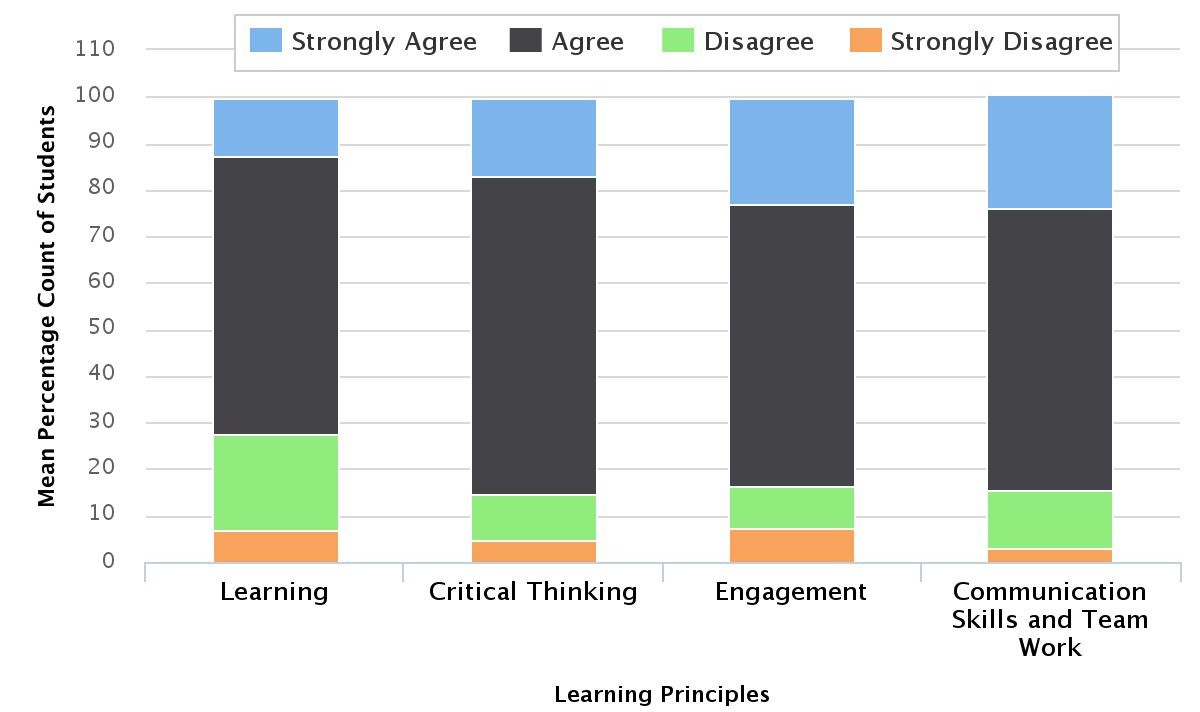}
\caption{Evaluation Results based on Course Survey by Students.}
\label{survey2}
\end{figure*}
\begin{figure*}
\centering
\noindent\includegraphics[width=14cm,height=7cm]{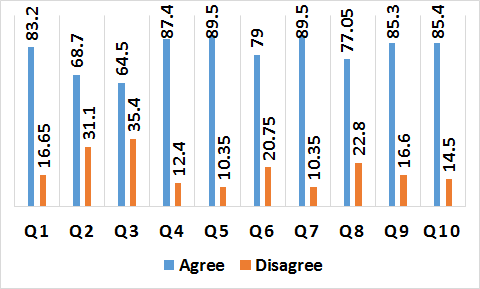}
\caption{Percentage of Agree and Disagree Responses for Ten Questions}
\label{q10}
\end{figure*}
\vspace{5 mm}
This case was looked by different perspective by different teams and thus they identified and synthesized different challenges. Each team gave different set of grammars stating different situations they could think can come into human-robot chess play. While some teams presented a very abstract view of the system in their grammar, few teams did incorporated detailed requirements of the system in their grammar. First the teams identified the tokens in the system. Some worked with taking tokens as chess states (like checkmate, draw), some worked with robot's movement (like set, release) as tokens while some used sensor's readings (like pressure release, pressure set) as tokens. 
\par
\setlength{\parindent}{5ex}
Few teams presented the view of making it essential for humans and robot to operate in their own workspace like waiting of the robot's arm to finish human's move. Some included the productions of the details of robot's behavior in handling chess pieces like touching, holding, sliding and reacquiring pieces when they fall. A few did work on providing equal chances to both players and ending the game on either a human win, robot win or a draw. Some of the team focused on working on the mechanics of the robot's arm like stretching, turning and grasping the piece. Students also worked on the grammar structure to reset the board like making space on the home square if some piece is already occupied. A few of the teams worked on including productions for different chess strategies like for en passant move taking the captured pawn and moving its own pawn to destination square. Some presented semantic actions also with the grammar. Teams also build motion parser using different bottom up parsing techniques discussed in the class. Thus by this case, students gained insights on how to create grammar for a real software.

\section{Evaluation}
\subsection{Experimental Setup}
The experiment is carried on compiler design course offered to third year B.Tech students of Deen Dayal Upadhyaya\footnote{http://dducollegedu.ac.in/} college, affiliated to Delhi University. There were about $48$ students in the class. We conduct students survey to record their responses to case-based teaching methodology. We analyze the survey using frequency distribution and chi square test of association to analyze if students agree that case discussion enhanced different learning principles like learning, critical thinking, engagement, communication skills and team work \cite{JEE:JEE1042}. The survey is adapted from a national survey on faculty perceptions of benefits and challenges of case-based instruction \cite{yadav2007teaching}. The questions in survey are changed to reflect students perspective on influence of case discussion on different learning subscales \cite{JEE:JEE1042}. Each question offers $4$ choices: Strongly Agree, Agree, Disagree and Strongly Disagree. Students choose one of the given option based on their experience about the case discussions.

\subsection{Effectiveness of Case-Based Learning}
Survey questions categorised under different learning principles along with percentage of students choosing the given option is shown in Table \ref{survey}. From Table \ref{survey} it can be observed that a high percentage of students $83.2\%$ (both agree and strongly agree) agreed that case discussions are relevant in learning about compiler concepts which shows that cases were effective in understanding vital components of compiler. About $68.7\%$ admitted that case discussions provided them deeper understanding of the concepts of compilers and thus enhanced their understanding. Relatively less but still significant percentage of students, $64.5\%$ were of the opinion that cases would help them to retain different aspects of compilers. Creating more cases for other phases of compilers can help for better overall retainment of compiler concepts. Students also felt that the use of case discussions enhanced their critical thinking. Specifically about $87.4\%$ thought case discussions gave them a good practice to view an issue from multiple perspectives. Thus it shows that cases developed are challenging and thought provoking. A significant percentage of students $89.5\%$ thought that case study was helpful in synthesizing ideas and information presented in course proving that cases covered the essential concepts of lexical and syntax analysis phase. Since real life problems were challenged in the cases, about $79\%$ students felt it added realism in the class. A majority of students $89.5\%$ agreed that they were more engaged in class during case discussions, showing that cases are interesting and engrossing. Formed cases are shown appropriate to compiler course as $77.05\%$ students reported that case discussions did increase their interest in learning about compilers. Students also felt that case discussions contributed to strengthen their communication skills ($85.3\%$) and it also increased their confidence to work in a team ($85.4\%$) which shows that case discussions helped in development of personal and interpersonal skills.  

Overall results show that students have a positive attitude towards the case-based learning methodology. The mean percentage of count of students along with their response to survey questions belonging to a given learning principle is computed and shown in Figure \ref{survey2}. From Figure \ref{survey2} it can be determined that
on an average $19.1\%$ strongly agreed to have achieved all the  learning principles during the case discussions. A significant percentage of students about $62.2\%$ on an average admitted to have acquired the skills of learning, critical thinking, engagement, communication skills and team work. A small percentage $13\%$ (mean) disagrees and yet another very small mean percentage of $5.4\%$ strongly disagrees to have acquired the learning principles during the process. The frequency of individual survey questions is aggregated to give each of the learning principles- learning, critical thinking, engagement, communication skills and team work a total frequency score and chi-square analysis is done over it. Values of ${\chi}^2 ($2$)=38.7$ and \textit{p}=.000013 suggests that there is relationship and association between different learning principles. Thus significantly more students agree that case studies increase the skills of learning, critical thinking, engagement, communication and team work. Figure \ref{q10} reveals the comparison between the agree (combining strongly agree and agree) and disagree (combining strongly disagree and disagree) responses for all the ten questions. The graph shows significant differences between the two bars and demonstrate the effectiveness of the approach based on student responses. 

\section{Challenges, Limitations and Recommendations}
As mentioned in previous literature \cite{mostert2007challenges} and observed by us also, there were various challenges that detracted from effective case-based teaching. Students were exposed to case-based teaching pedagogy for the first time. They were used to the standard teaching environment of listening to the lectures, taking notes and interacting with teachers only during questions and answers. Thus for them adjusting to an unfamiliar environment where they are required to apply their knowledge to complex ambiguous problems, to reflect, discuss and debate was a little difficult. Relating case to the theoretical contents did not come naturally to them. Undergraduates did not have practical experience to case-based discussions and some of them viewed practical aspects of cases different from theoretical ones. We also experienced that case preparation is a laborious activity. We had to contemplate the purpose for which the case is used and the course content to which it is linked while developing the case. Since case discussions rely heavily on discussions, participants ability to convey good communication is essential. We observed that in the class some of the students were hesitant to participate and express their views. 
 \par
\setlength{\parindent}{5ex}
We also observe some limitations of case discussions as it has been stated in literature too \cite{INSP:INSP002}. Case discussions consume a lot of time thus managing time lectures with cases should be done well. Finding the correct cases which covers all the essential learning objectives is challenging. Since cases have multiple perspective, it may happen that a student interprets it in completely different manner than what the instructor wants. This might lead to missing of the learning aspects desired by the intructor. Grading is subjective as there is no completely right or wrong answer. There are more then one way to look at the case, thus there is problem in validation of solutions presented by students. Also since students are working in the group, grade is to assigned to the group itself thus it is hard to determine individual contribution of each team member. 
 \par
\setlength{\parindent}{5ex}
We can suggest some recommendations based on our personal experiences. It is challenging for a single instructor to manage and facilitate several groups. The instructor and group interaction is paramount, without it the entire purpose of learning through discussions will be defeated. Thus we recommend to use a additional teaching support for better outcome. To make group discussion easier, it is better to use flip-charts, markers and round tables. Just like there are case repositories for science subjects\footnote{http://sciencecases.lib.buffalo.edu/cs/collection/websites.asp}, business management\footnote{http://www.ibscdc.org/} we recommend to create a central repository of compiler design cases so that multiple instructors all over the world can benefit.

\section{Conclusion}
We propose and successfully implement the CPLC teaching environment for teaching compiler design concepts using case-based and project-based pedagogy. With case-based pedagogy students gave positive feedback to have learned the course, developed skill of critical thinking about an issue, being actively involved in the course and having improved communication skills and team work. The hands on experience on project gave more practical experience of designing the compiler by themselves. The positive measurement of effectiveness of case-based discussion along with practical experience of compiler designing through projects proves that both of these pedagogy are suitable for teaching concepts of compiler. 

\bibliographystyle{IEEEtran}  
\bibliography{T4E}  
\end{document}